\documentclass[aps,pre,onecolumn]{revtex4}

\usepackage{bm}
\usepackage{graphicx}
\usepackage{amsmath}
\usepackage{amssymb}

\newcommand{\Chapo}[1]{\chi_{_{\!#1}}}
\newcommand{\uu}{u^\ulcorner}
\newcommand{\tu}{t^\ulcorner}
\newcommand{\ul}{u_\lrcorner}
\newcommand{\tl}{t_\lrcorner}
\newcommand{\U}{\mathcal{U}}
\newcommand{\T}{\mathcal{T}}

\DeclareMathOperator*{\diag}{diag}

\begin{document}

\title{Diagonalization of quasi-uniform tridiagonal matrices}

\author{Leonardo Banchi}
\affiliation{ISI Foundation, Via Alassio 11/c,
        I-10126 Torino (TO), Italy}

\author{Ruggero Vaia}
\affiliation{Istituto dei Sistemi Complessi,
             Consiglio Nazionale delle Ricerche,
             via Madonna del Piano 10,
             I-50019 Sesto Fiorentino (FI), Italy}

\date{\today}

\begin{abstract}
The task of analytically diagonalizing a tridiagonal matrix can be
considerably simplified when a part of the matrix is uniform. Such
quasi-uniform matrices occur in several physical contexts, both
classical and quantum, where one-dimensional interactions prevail.
These include magnetic chains, 1D arrays of Josephson junctions or of
quantum dots, boson and fermion hopping models, random walks, and so
on. In such systems the bulk interactions are uniform, and
differences may occur around the boundaries of the arrays. Since in
the uniform case the spectrum consists of a band, we exploit the bulk
uniformity of quasi-uniform tridiagonal matrices in order to express
the spectral problem in terms of a variation of the distribution of
eigenvalues in the band and of the corresponding eigenvectors. In the
limit of large matrices this naturally leads to a deformation of the
density of states which can be expressed analytically; a few
out-of-band eigenvalues can show up and have to be accounted for
separately. The general procedure is illustrated with some examples.
\end{abstract}


\maketitle

\section{Introduction}
\label{s.intro}

The diagonalization of quasi-uniform tridiagonal matrices, namely
tridiagonal matrices which are uniform except at the boundaries,
appears in many branches of mathematics and
physics~\cite{Yueh2005,DaFonseca2007,kouachi2006eigenvalues,%
Thomas1995,Mulken2011}. In particular, tridiagonal matrices generally
occur in the theory of one-dimensional lattices with
nearest-neighbour interactions. In this context quasi-uniform
tridiagonal matrices have been recently applied for achieving high
quality quantum communication between distant
parts~\cite{banchi2011long,apollaro2006entanglement,%
Banchi2011a,Yao2011,Wojcik2005,GiampaoloI2010}, and for describing
spin systems in a spin
environment~\cite{Apollaro2010,Tjon1970,Stolze2000}.

In this paper we put forward a general method for calculating the
eigenvalues and the eigenvectors of symmetric tridiagonal matrices by
exploiting the property of bulk uniformity. This allows us to put the
eigenvalues in the form of deformations, defined by suitable {\em
shifts}, of those of the fully uniform case, which are known to form
a band. The modified density of the eigenmodes in the band is
expressed in terms of functions which can be analytically evaluated
and depend on the non-uniform matrix elements. A small number of
localized eigenstates could emerge from the band and have to be
accounted for separately. Particular applications of this technique
were made in Refs.~\cite{banchi2011long,apollaro201299}.

\section{Tridiagonal matrices}
\label{s.thmtrig}

A symmetric $\ell\times\ell$ tridiagonal matrix $T=\{T_{\mu\nu}\}$
has $2\ell{-}1$ independent real elements, namely
$T_{\mu\mu}\equiv{a_\mu}$ ($\mu\,{=}\,1,\dots,\ell$) and
$T_{\mu,\mu+1}=T_{\mu+1,\mu}\equiv{b_\mu}$ ($\mu=1,\dots,\ell{-}1$).
Its spectral decomposition is $T=O^\dagger\,\Lambda\,O$, where
$O=\{O_{k\mu}\}$ is orthogonal, its rows being the $\ell$
eigenvectors of $T$ with eigenvalues $\lambda_k$, and
$\Lambda=\diag(\{\lambda_k\})$.

$T$ is said to be {\em mirror-symmetric} if it is also symmetric with
respect to the skew diagonal, namely $[T,J]=0$~, where
$J_{\mu\nu}=\delta_{\mu,\ell{+}1{-}\nu}$ is the mirroring matrix. In
the mathematical language such matrices are both {\em persymmetric}
($JTJ=T^t$) and {\em centrosymmetric} ($JTJ=T$). It is known that the
eigenvectors of a mirror-symmetric $T$ are either symmetric or
antisymmetric~\cite{CantoniB1976},
\begin{equation}
  O_{k,\ell{+}1-\mu} = (-)^{k+1} O_{k\mu}~;
  \label{e.cantoni}
\end{equation}
this formula assumes that $b_\mu>0$ and the eigenvalues
$\{\lambda_k\}$ are listed in decreasing order.

The eigenvectors can be completely expressed in terms of
characteristic polynomials of submatrices of $T$, evaluated at the
eigenvalues. In order to prove this, let us introduce the following
notation for tridiagonal submatrices,
\begin{equation}
  T_{\mu:\nu} = \begin{pmatrix}
    a_\mu & b_\mu & \\
    b_\mu& a_{\mu+1}&b_{\mu+1} \\
    &b_{\mu+1} & \ddots & \ddots \\
    &&\ddots & \ddots \\
    &&&&a_{\nu-1} & b_{\nu-1} \\
    &&&&b_{\nu-1} & a_{\nu} \\
  \end{pmatrix}~,
  \label{e.tridiag}
\end{equation}
and for the corresponding characteristic polynomials,
\begin{equation}
 \Chapo{\mu:\nu}(\lambda)=\det[\lambda-T_{\mu:\nu}] ~,
\end{equation}
where $\mu\,{\le}\,\nu$; then $T_{1:\ell}\equiv{T}$ and
$\chi_{1:\ell}(\lambda)\equiv{\chi(\lambda)}$; the eigenvalues are
the $\ell$ solutions of the secular equation
$\chi(\lambda_k)\,{=}\,0$. By expanding from the bottom (upper)
corners, these polynomials are found to satisfy the recurrence
relations
\begin{subequations}
  \begin{align}
    \Chapo{\mu:\nu}(\lambda)& = (\lambda-a_\nu)\;
    \Chapo{\mu:\nu-1}(\lambda) -
    b_{\nu-1}^2\; \Chapo{\mu:\nu-2}(\lambda) ,
   \label{e.trirecurra}
 \\
    \Chapo{\mu:\nu}(\lambda)& = (\lambda-a_\mu)\;
    \Chapo{\mu+1:\nu}(\lambda) -
    b_\mu^2\; \Chapo{\mu+2:\nu}(\lambda)~.
  \label{e.trirecurrb}
  \end{align}
  \label{e.trirecurr}
\end{subequations}

The following important and useful formula (see, e.g.,
Ref.~\cite{Parlett1998}) expresses the product of two components of
the same eigenvector,
\begin{equation}
  \chi'(\lambda_k) \;O_{k\mu}O_{k\nu} =
  \Chapo{1:\mu-1}(\lambda_k)\;\bigg(\prod_{\sigma=\mu}^{\nu-1}b_\sigma\bigg)\;
  \Chapo{\nu+1:\ell}(\lambda_k) ~,
  \label{e.paigethm}
\end{equation}
which holds for $\mu\le\nu$ if one defines
$\Chapo{1:0}(\lambda_k)\,{=}\,\Chapo{\ell+1:\ell}(\lambda_k)\,{\equiv}\,1$.
One can assume $b_\mu\neq{0}$ for all $\mu=1,\dots,\ell{-1}$, as
otherwise the diagonalization of $T$ would split into the
diagonalization of independent submatrices, so that the eigenvalues
of $T$ are nondegenerate. Hence, the derivatives of the
characteristic polynomial at the eigenvalues do not vanish,
$\chi'(\lambda_k)\neq{0}$, and Eq.~~\eqref{e.paigethm} can be solved
for the eigenvector components, for example
\begin{equation}
 O_{k1}^2 =  \frac{\Chapo{2:\ell}(\lambda_k)}{\chi'(\lambda_k)} ~,
~~~~~~~~
 O_{k\ell}^2=\frac{\Chapo{1:\ell-1}(\lambda_k)}{\chi'(\lambda_k)}~,
  \label{e.Okboundary}
\end{equation}
and from one of these (one can arbitrarily choose the positive root)
the remaining elements of the $k$-th eigenvector follow by means of
Eq.~\eqref{e.paigethm}; for instance, taking $\mu\,{=}\,1$,
\begin{equation}
  O_{k\nu} = O_{k1}~ b_1\cdots b_{\nu-1}
  \frac{\Chapo{\nu+1:\ell}(\lambda_k)}{\Chapo{2:\ell}(\lambda_k)}
  = O_{k1}~\prod_{\mu=2}^\nu b_{\mu-1}
  \frac{\Chapo{\mu+1:\ell}(\lambda_k)}{\Chapo{\mu:\ell}(\lambda_k)}~.
  \label{e.Oknu}
\end{equation}
This shows that the orthogonal matrix $O$ can be fully expressed in
terms of characteristic polynomials.

Note also that the recurrence equations~\eqref{e.trirecurr} give
\begin{equation}
 O_{k,\nu+1} = \frac{\lambda_k-a_\nu}{b_\nu}\, O_{k\nu}
         -\frac{b_{\nu-1}}{b_\nu}\, O_{k,\nu-1}
         ~~~~~(\nu=1,\dots,\ell{-}1)~,
 \label{e.evecrecur}
\end{equation}
with the assumption $O_{k0}\,{=}\,0$; these equations can be used for
a sequential computation of the eigenvectors' components once the
eigenvalues are known. An important consequence of this construction
is that, once the first components of the eigenvectors ($O_{k1}$) are
determined from Eq.~\eqref{e.Okboundary}, the eigenvectors come out
already normalized, i.e., the matrix $O$ is orthogonal, making the
explicit normalization unnecessary and tremendously simplifying the
analytical calculations.

When the matrix size $\ell$ is large the characteristic polynomials
$\Chapo{\mu:\nu}(\lambda_k)$ have a high degree and the analytical
evaluation of the eigenvalue decomposition is very demanding. In the
forthcoming section we provide general simplified formulas for the
eigenvalues and for the eigenvector elements,
Eqs.~\eqref{e.Okboundary} and~\eqref{e.Oknu}, in the case of a
quasi-uniform matrix $T$.

\section{Quasi-uniform tridiagonal matrices}%
\label{s.trigdiag}

\subsection{Uniform tridiagonal matrices}

A uniform tridiagonal matrix has equal elements within each diagonal,
namely $a_\mu=a$ and $b_\mu=b$, and without loss of generality one
can set $b=1$ and $a=0$. In this case the recurrence
relations~\eqref{e.trirecurr} for the characteristic polynomials are
found to be equal to those defining the Chebyshev polynomials of the
second kind~\cite{abramowitz1964handbook},
\begin{equation}
  \U_n(\xi)
  = \frac{\big(\xi+\sqrt{\xi^2{-}1}\big)^{n+1}-
          \big(\xi-\sqrt{\xi^2{-}1}\big)^{n+1}}
          {2\sqrt{\xi^2{-}1}}~,
\label{e.chebyshev}
\end{equation}
the correspondence being
$\Chapo{1:\ell}(\lambda)=\U_\ell(\lambda/2)$. Setting
$\xi\equiv\cos{k}$ the Chebyshev polynomials of the second kind can
be compactly written as
\begin{equation}
 \U_n(\cos{k})=\frac{\sin[(n{+}1)k]}{\sin{k}} ~,
\label{e.Ucosk}
\end{equation}
so that the secular equation $\chi(\lambda)=\U_\ell(\lambda/2)=0$
defines the $\ell$ eigenvalues $\lambda\equiv2\cos{k}$ corresponding
to
\begin{equation}
 k \equiv k_j = \frac{\pi j}{\ell+1}~, ~~~~~~~~(j=1,\dots,\ell) ~.
\label{e.kjunif}
\end{equation}
With no ambiguity we will use henceforth the index $k$ as running
over such a set of $\ell$ discrete values, so we may keep the
notations introduced for the spectral decomposition and, e.g., write
the eigenvectors of the uniform case as
$O_{k\mu}=\sqrt{2/(\ell{+}1)}\,\sin(\mu{k})$.

\subsection{Quasi-uniform tridiagonal matrices}

A tridiagonal matrix $T$ is said to be {\em quasi uniform} if it is
mainly constituted by a large uniform tridiagonal block $T_{u:v}$ of
size $n{\times}n$ (with $n\,{=}\,v{-}u{+}1$), i.e., its elements are
$a_u=a_{u+1}=\dots=a_v\equiv{a}$ and
$b_u=b_{u+1}=\dots=b_{v-1}\equiv{b}$. By `large uniform block' it is
meant that the number of different elements, sitting at one or both
{\em corners}, is much smaller than the size of the whole matrix $T$,
namely that $\ell{-}n\ll\ell$. Indeed, the important point of our
approach is in taking into account the uniform part of $T$, which for
quasi-uniform tridiagonal (QUT) matrices is almost the whole $T$, and
use the properties of Chebyshev polynomials for reducing the
complexity of Eqs.~\eqref{e.paigethm} and~\eqref{e.Okboundary}.
Again, without loss of generality we set $a=0$ and $b=1$ in what
follows.

The results we present in this paper are based on the following
important statement: the characteristic polynomial of QUT matrices
can be expressed in terms of the Chebyshev
polynomials~\cite{abramowitz1964handbook} of the first and second
kind, $\T_{n+1}(\xi)$ and $\U_{n}(\xi)$,
\begin{equation}
  \chi(2\xi) \equiv \Chapo{1:\ell}(2\xi) =
  u(\xi) \,\U_{n}(\xi)+t(\xi)\,
  \T_{n+1}(\xi)~,
  \label{e.chebydec}
\end{equation}
where $u(\xi)$ and $t(\xi)$ are {\em low-degree} polynomials: indeed,
their degree cannot be larger than $\ell{-}n$ and $\ell{-}n{-}1$,
respectively. Their coefficients involve the nonuniform matrix
elements and generally they can be easily calculated by means of
Eqs.~\eqref{e.trirecurr}.

In order to prove the above general statement we start from the
characteristic polynomial of the uniform tridiagonal submatrix
$T_{u:v}$ and calculate the characteristic polynomial of larger
submatrices by means of Eq.~\eqref{e.trirecurra}:
\begin{align}
  \Chapo{u:v}(2\xi) & =  \U_n(\xi)
\nonumber \\
  \Chapo{u:v+1}(2\xi) & =  (2\xi-a_{v+1})\,\U_n(\xi) -
  b_v^2 \,\U_{n-1}(\xi)
\nonumber \\
  \Chapo{u:v+2}(2\xi) & =  (2\xi-a_{v+2})\,\Chapo{u:v+1}(2\xi)-
  b_{v+1}^2 \,\U_{n}(\xi)
\nonumber \\
  & \quad \vdots
\nonumber \\
  \Chapo{u:\ell}(2\xi) & = \tilde p_0(\xi) \,\U_n(\xi)
  + \tilde p_1(\xi)\,\U_{n-1}(\xi) ~;
\label{e.chiul}
\end{align}
this holds for some polynomials
$\tilde{p}_0(\xi)=(2\xi)^{\ell{-}v}+\dots$ and $\tilde{p}_1(\xi)$,
whose coefficients are products of the nonuniform matrix elements
$a_{v+1},\dots,a_\ell$ and $b_\nu,\dots,b_\ell$. By further enlarging
the matrix $T_{u:\ell}$ in the upper corner by means of
Eq.~\eqref{e.trirecurrb} we first obtain
\[
  \Chapo{u-1:\ell}(2\xi) = (2\xi - a_{u-1})\,
  \Chapo{u:\ell}(2\xi) - b_{u-1}^2\,
  \Chapo{u+1:\ell}(2\xi)~,
\]
where $\Chapo{u+1:\ell}(2\xi)$ concerns the QUT matrix $T_{u+1:\ell}$
whose uniform block is $(n{-}1){\times}(n{-}1)$, so its expression
analogous to Eq.~\eqref{e.chiul} involves $\U_{n-1}(\xi)$ and
$\U_{n-2}(\xi)$. Proceeding further one has
\begin{align}
  \Chapo{u-2:\ell}(2\xi) &= (2\xi - a_{u-2})\,
  \Chapo{u-1:\ell}(2\xi) - b_{u-2}^2\,
  \Chapo{u:\ell}(2\xi)
\nonumber \\
  & \quad \vdots
\nonumber \\
  \Chapo{1:\ell}(2\xi) & = p_0(\xi)\,\U_n(\xi)
    + p_1(\xi)\,\U_{n-1}(\xi)+ p_2(\xi)\,\U_{n-2}(\xi)~,
\label{e.chi1l}
\end{align}
for some polynomials $p_0(\xi)$, $p_1(\xi)$, and $p_2(\xi)$. This
expression allows us to recover Eq.~\eqref{e.chebydec}, using the
identities
\begin{subequations}\label{e.chebprop1}
\begin{align}
  \U_{n-1}(\xi) & =  \xi \, \U_{n}(\xi) -
  \T_{n+1}(\xi)~,
\\
  \U_{n-2}(\xi) & =2\xi\,\U_{n-1}(\xi)-\U_n(\xi)
                  =(2\xi^2{-}1)\,\U_{n}(\xi)-2\xi\T_{n+1}(\xi)~,
\end{align}
\end{subequations}
and identifying
\begin{subequations}
\begin{align}
 u(\xi) &= p_0(\xi)+\xi\,p_1(\xi)+(2\xi^2-1)p_2(\xi)~,
\\
 t(\xi) &= -p_1(\xi)-2\xi\,p_2(\xi)~.
\end{align}
\end{subequations}

The usefulness of expressing
$\chi(\lambda)\equiv\Chapo{1:\ell}(\lambda)$ in the
form~\eqref{e.chebydec} is evident looking at the analog of
Eq.~\eqref{e.Ucosk} for the first-kind Chebyshev polynomials,
\begin{equation}
 \T_n(\cos{k})= \cos(nk)~,
\label{e.Tcosk}
\end{equation}
which turns Eq.~\eqref{e.chebydec} into
\begin{equation}
 \chi(2\cos{k}) =
 u(\cos{k})\,\frac{\sin[(n{+}1)k]}{\sin{k}}+t(\cos{k})\,\cos[(n{+}1)k]~,
\end{equation}
hence, the secular equation $\chi\,{=}\,0$ can be written
\begin{equation}
 \sin[(n{+}1)k-2\phi_k]=0~,
\label{e.secularkn}
\end{equation}
with the angle $\phi_k$ defined by
\begin{equation}
 \tan{2\phi_k}=-\frac{\sin{k}~t(\cos{k})}{u(\cos{k})}~.
\label{e.varphik}
\end{equation}
Equivalently, the same form of the secular equation can be derived
directly by simply rewriting Eq.~\eqref{e.Ucosk} as
~$\sin{k}\,\U_n(\cos{k})=\Im\big\{e^{i(n{+}1)k}\big\}$~ and replacing
it in Eq.~\eqref{e.chi1l}, which turns indeed into
\begin{equation}
 \Im\big\{e^{i[(n{+}1)k-2\phi_k]} \big\} = 0 ~,
\end{equation}
where $2\phi_k$ coincides with the phase of the complex number
\begin{equation}
 w_k \equiv p_0(\xi)+ e^{-ik}\,p_1(\xi)+ e^{-2ik}\,p_2(\xi)
     = |w_k|\,e^{-2i\phi_k}~.
\label{e.wkphik}
\end{equation}

It is convenient, in order to easily recover the limit of a fully
uniform $\ell{\times}\ell$ matrix $T$, to use slightly modified
versions of Eqs.~\eqref{e.secularkn} and~\eqref{e.varphik}, namely
\begin{equation}
  \sin[(\ell{+}1)k-2\varphi_k] = 0~,
\label{e.seculark}
\end{equation}
with \emph{shifts} $\varphi_k$ defined by
\begin{equation}
 2\varphi_k = (\ell{-}n)\,k
              - \tan^{-1}\frac{t(\cos{k})\,\sin k}{u(\cos{k})}~.
\label{e.phik}
\end{equation}

Hence, the eigenvalues of the QUT matrix, parametrized as
$\lambda=2\cos{k}$, with $k\in[0,\pi]$, can be obtained from the
equations
\begin{equation}
 k \equiv k_j = \frac{\pi\,j+2\varphi_{k_j}}{\ell{+}1}
 ~,~~~~~ (j=1,\dots,\ell)~,
\label{e.kj}
\end{equation}
which determine the {\em allowed} values of $k$. Comparing with
Eq.~\eqref{e.kjunif} it appears that the shifts $\varphi_k$ represent
the deviation from the uniform case, where they vanish.
Eq.~\eqref{e.kj} can be solved numerically for any $j$ (except for a
few $j$'s if there are out-of-band eigenvalues, see below). Usually,
an iterative computation is fast converging; in the limit of
$\ell\gg{1}$ even the truncation of~\eqref{e.kj} after the first
iteration can be very accurate, as it was verified in the cases
considered in Refs.~\cite{banchi2011long,apollaro201299}.

Noteworthy, in the limit of large $\ell$ Eq.~\eqref{e.kj} allows us
to obtain a useful analytic expression of the {\em density of states}
$\rho_k$ defined in the interval $k\in[0,\pi]$, namely
\begin{equation}
 \rho_k^{-1} = \partial_j k = \frac{\pi}{\ell{+}1-2\varphi_k'} ~,
\label{e.dos}
\end{equation}
by means of which summations over eigenmodes can be transformed into
integrals over $k$,
\begin{equation}
 \sum_j ~(\cdots) ~~\simeq~~  \int_0^\pi dk~\rho_k ~(\cdots) ~;
\label{e.sumintegral}
\end{equation}
one can also observe that $\rho_k^{-1}$ represents the spacing
between subsequent allowed values of $k$: the deformation from the
equally-spaced $k$'s of the uniform case, $\pi/(\ell{+}1)$, is
represented by the correction term with $\varphi_k'$.

\subsection{Out-of-band eigenvalues}
\label{ss.oob}

The fact of setting $\lambda\,{\equiv}\,2\cos{k}$ does not imply that
{\em all} eigenvalues are included in the band $[-2,2]$. For a QUT
matrix this is generally true for the largest part of the spectrum,
though a few eigenvalues can emerge over or below the band when (the
absolute values of) the nonuniform matrix elements are large enough;
correspondingly, Eq.~\eqref{e.kj} cannot be solved for a few values
of $j$, i.e., Eq.~\eqref{e.seculark} has less than $\ell$ solutions
in the interval $k\in[0,\pi]$. On the other hand, the out-of-band
eigenvalues are still described by $\lambda\,{\equiv}\,2\cos{k}$, but
with complex values of $k=q\,{+}\,ip$; for the eigenvalues to be real
$q$ must be either $0$ or $\pi$, i.e.,
\begin{equation}
 \lambda\,{=}\,\pm2\cosh{p}~,
\label{e.lambdach}
\end{equation}
and $p\ge{0}$. Correspondingly, one can take the expression for the
Chebyshev polynomials when the absolute value of the argument is
larger than one,
\begin{equation}
 U_{n}(\pm\cosh{p}) = (\pm)^n~\frac{\sinh{(n{+}1)p}}{\sinh{p}}~.
\label{e.Uch}
\end{equation}
In the large-$\ell$ limit, the out-of-band eigenvalues have to be
considered separately by adding to the integral~\eqref{e.sumintegral}
the sum over the out-of-band states. An example of how to deal with
such eigenvalues is given in Section~\ref{ss.wurst}.

\subsection{Eigenvectors}

The boundary elements of the eigenvectors given in
Eq.~\eqref{e.Okboundary} can be calculated using the same formalism.
Indeed, following the construction of the previous subsection we can
find the polynomials $\uu(\xi)$, $\tu(\xi)$, $\ul(\xi)$, $\tl(\xi)$
such that
\begin{align}
  \Chapo{2:\ell}(2\xi) =  \uu(\xi)\,\U_{n}(\xi)
                         +\tu(\xi)\,\T_{n+1}(\xi)~,
\notag\\
  \Chapo{1:\ell-1}(2\xi)= \ul(\xi)\,\U_{n}(\xi)
                         +\tl(\xi)\,\T_{n+1}(\xi)~,
\label{e.utcorner}
\end{align}
where the symbols $\ulcorner$ and $\lrcorner$ clearly refer to the
submatrices  $T_{2:\ell}$ and $T_{1:\ell{-}1}$, respectively.
Accordingly, expressing $\chi'(\lambda)$ as a function of $\U_n$ and
$\T_{n+1}$ thanks to the relations
\begin{subequations}\label{e.chebprop2}
\begin{align}
 \T'_{n+1}(\xi)
   &= (n{+}1)~\U_{n}(\xi)~,
\\
 (1{-}\xi^2)~\U'_{n}(\xi)
   &= \xi~\U_{n}(\xi)- (n{+}1)\,\T_{n+1}(\xi) ~,
\end{align}
\end{subequations}
Eqs.~\eqref{e.Okboundary} take the form
\begin{subequations}
\begin{align}
  O_{1k}^2 &= 2 \frac{\uu(\xi_k)\, \U_{n}(\xi_k)+
  \tu(\xi_k)\, \T_{n+1}(\xi_k)}{
  u^\star_n(\xi_k)\, \U_{n}(\xi_k)+t^\star_n(\xi_k)\,
  \T_{n+1}(\xi_k)}~,\\
  O_{\ell k}^2 &= 2 \frac{\ul(\xi_k)\, \U_{n}(\xi_k)+
  \tl(\xi_k)\, \T_{n+1}(\xi_k)}{
  u^\star_n(\xi_k)\, \U_{n}(\xi_k)+t^\star_n(\xi_k)\,
  \T_{n+1}(\xi_k)}~,
\end{align}
  \label{e.evxx1boundche}
\end{subequations}
where $\xi_k\equiv\lambda_k/2\equiv\cos{k}$ and
\begin{subequations}\label{e.chebcoeffprim}
  \begin{align}
    u^\star_n(\xi) & =  u'(\xi)+\frac{\xi }{1{-}\xi^2}\,u(\xi)+
    (n{+}1)\, t(\xi), \\
    t^\star_n(\xi) & =  t'(\xi)-\frac{n{+}1}{1{-}\xi^2}\,u(\xi).
\end{align}
\end{subequations}
As the eigenvalues are the solutions of the secular equation,
\begin{equation}
  0 = u(\xi_k) \,\U_{n}(\xi_k)+t(\xi_k)\,
  \T_{n+1}(\xi_k)~,
  \label{e.seculareqc}
\end{equation}
the high-degree polynomials $\U_{n}(\xi_k)$ and $\T_{n+1}(\xi_k)$ can
be removed from~\eqref{e.evxx1boundche} and accordingly
\begin{subequations}
  \label{e.ok1formal}
  \begin{align}
    O_{1k}^2 &= 2 \frac{\uu(\xi _k)\, t(\xi _k)-
    \tu(\xi _k)\, u(\xi _k)}{
    u^\star_n(\xi _k)\, t(\xi _k)-
    t^\star_n(\xi _k)\, u(\xi _k)}~,
    \\
    O_{\ell k}^2 &= 2 \frac{\ul(\xi _k)\, t(\xi _k)-
    \tl(\xi _k)\, u(\xi _k)}{
    u^\star_n(\xi _k)\, t(\xi _k)-t^
    \star_n(\xi _k)\, u(\xi _k)}~.
  \end{align}
\end{subequations}
This shows a remarkable result, namely that, although the eigenvector
components generally depend on complicated high-degree polynomials,
for QUT matrices one can express the boundary coefficients of the
eigenvectors in terms of ratios of low-degree polynomials.

Further simplifications can be obtained by replacing again
$\xi_k\,{=}\,\cos{k}$. In fact, from Eq.~\eqref{e.phik}
\begin{equation}
 2\varphi_k'= (\ell{-}n)
  -\frac{tu\,\cos k +(t'u-u't)\,\sin^2 k}{u^2+t^2\sin^2k}~,
\end{equation}
where the argument $\xi_k$ of $u$ and $t$ is understood, so that the
eigenvector elements~\eqref{e.ok1formal} read
\begin{subequations}\label{e.ok1cosk}
\begin{align}
  O_{1k}^2 &= \frac{2\sin^2k}{\ell{+}1-2\varphi_k'}\;
 \frac{\uu(\xi_k)\, t(\xi_k)-
\tu(\xi_k)\, u(\xi_k)}{
u^2(\xi_k)+t^2(\xi_k)\;\sin^2k}~,
\\
O_{\ell k}^2 &= \frac{2\sin^2k}{\ell{+}1-2\varphi_k'}\;
\frac{\ul(\xi_k)\, t(\xi_k)-
\tl(\xi_k)\, u(\xi_k)}{
u^2(\xi_k)+t^2(\xi_k)\;\sin^2k}~.
\end{align}
\end{subequations}
These expressions generalize what was found in
Refs.~\cite{banchi2011long,apollaro201299}.

As for the remaining elements, note that the recurrence
relation~\eqref{e.evecrecur} in the bulk, i.e., for $u<\nu<v$, reads
\begin{equation}
 O_{k,\nu+1}+O_{k,\nu-1}=(e^{ik}+e^{-ik})\,O_{k\nu} ~,
\end{equation}
whose generic solution is
\begin{equation}
 O_{k\nu}=\Im\{e^{ik\nu}\alpha_k\}~,
\label{e.solgen.Oknu}
\end{equation}
for any complex number $\alpha_k$ independent of $\nu$, which has to
be determined by requiring that the `boundary'
relations~\eqref{e.evecrecur}, i.e., for $\nu=2,\dots,u$ and
$\nu=v,\dots,\ell{-}1$ be satisfied.

\section{Examples}

\subsection{Persymmetric two-edge matrix}
\label{ss.wurst}

As a first example we consider a mirror-symmetric QUT matrix with two
non-uniform edges: the uniform block is of size $n=\ell{-}2$, so the
matrix reads, setting $b=1$ and $a=0$,
\begin{equation}\label{e.trigmat1bound}
T=T_{1:\ell}=\begin{pmatrix}
     x & y &&&&&  \\
    y & 0 & 1 &&&&  \\
    & 1 & 0 & 1 &&&  \\
    && 1 &  \ddots &\ddots&  \\
    &&&\ddots&&1&\\
    &&&& 1&  0 & y  \\
    &&&&& y &  x
\end{pmatrix}~,
\end{equation}
and, with the notations of the previous section, $u=2$ and
$v=\ell{-}1$.

Keeping the notation $\lambda\equiv2\xi$, thanks to the recursion
relations~\eqref{e.trirecurr} it holds that
\begin{align}\label{e.ximinori}
\Chapo{2:\ell}(2\xi) & = (2\xi{-}x) \,\U_n(\xi)-y^2 \,\U_{n-1}(\xi)
\notag \\
\Chapo{3:\ell}(2\xi) & = (2\xi{-}x) \,\U_{n-1}(\xi)-y^2 \,\U_{n-2}(\xi)
\notag\\
 \Chapo{1:\ell}(2\xi) & = (2\xi{-}x)\,\Chapo{2:\ell}(2\xi) - y^2\Chapo{3:\ell}
\notag\\
                      & = (2\xi{-}x)^2 \,\U_{n}(\xi)
  - 2(2\xi{-}x)y^2\,\U_{n-1}(\xi) +y^4 \,\U_{n-2}(\xi)~.
\end{align}
Accordingly, the secular equation for the in-band eigenvalues is
given by~\eqref{e.seculark}, where the shifts are more easily found
from Eq.~\eqref{e.wkphik}: indeed, thanks to mirror symmetry, $w_k$
turns out to be a square,
\begin{equation}
  w_k = (2\xi-x-y^2e^{-ik})^2
      =\big[(2{-}y^2)\cos{k}-x+iy^2\sin{k}\big]^2~,
\end{equation}
so that
\begin{equation}
  \varphi_k = k- \tan^{-1} \frac{y^2 \sin k}{(2{-}y^2)\cos k-x}~.
  \label{e.1.phik}
\end{equation}
The expression~\eqref{e.ximinori} can be rewritten in the
form~\eqref{e.chebydec} by means of the
properties~\eqref{e.chebprop1}, so with the notation of the previous
section we identify the coefficients of Eqs.~\eqref{e.chebydec}
and~\eqref{e.utcorner} as
\begin{subequations}
\begin{align}
 u(\xi) & = \big[(2{-}y^2)\xi-x\big]^2-y^4\,(1-\xi^2)~, \\
 t(\xi) & = 2y^2\,\big[(2{-}y^2)\xi-x\big]~, \\
 \uu(\xi) & = \ul(\xi) = (2{-}y^2)\xi-x~, \\
 \tu(\xi) & = \tl(\xi) = y^2~.
\end{align}
\label{e.xx1boundcoeff}
\end{subequations}
Of course, Eq.~\eqref{e.1.phik} can be obtained using straightforward
trigonometric identities also from~\eqref{e.phik} and the above
polynomials. As for the first components of the eigenvectors, they
follow from Eq.~\eqref{e.ok1cosk}:
\begin{equation}
  {O}_{k1}^2 = O_{k\ell}^2 = \frac{2}{\ell{+}1{-}2\varphi_k'}
  ~\frac{y^2\sin^2\!{k}}
 {[(2{-}y^2)\cos{k}-x]^2+y^4\sin^2\!{k}}~.
\label{e.Parlett4}
\end{equation}
Moreover, imposing to the generic solution~\eqref{e.solgen.Oknu} the
conditions~\eqref{e.evecrecur} at the corners, one finds
\begin{equation}
 \alpha_k=\frac{1-x\,e^{-ik}+(1{-}y^2)\,e^{-2ik}}{y\sin{k}}~{O_{k1}}~,
\end{equation}
so that all the components of the eigenvectors have a fully
analytical expression:
\begin{align}
  O_{k\nu} = \frac{\sin \nu k - x \sin(\nu{-}1)k
            + (1{-}y^2)\sin(\nu{-}2)k}{y \sin k}~O_{k1} ~,
  \label{e.wurstvec}
\end{align}
for $\nu=2,\dots,\ell-1$. Eqs.~\eqref{e.wurstvec}
and~\eqref{e.Parlett4}, together with~\eqref{e.1.phik}
and~\eqref{e.seculark}, give a complete solution to the analytical
diagonalization problem of the matrix~\eqref{e.trigmat1bound}. Note
that for $x\,{=}\,0$ this expression is in agreement with
Ref.~\cite{Wojcik2005} and that the term with $\varphi_k'$ becomes
irrelevant for large $n$.

\begin{figure}
\centering
\includegraphics[width=100mm,angle=0]{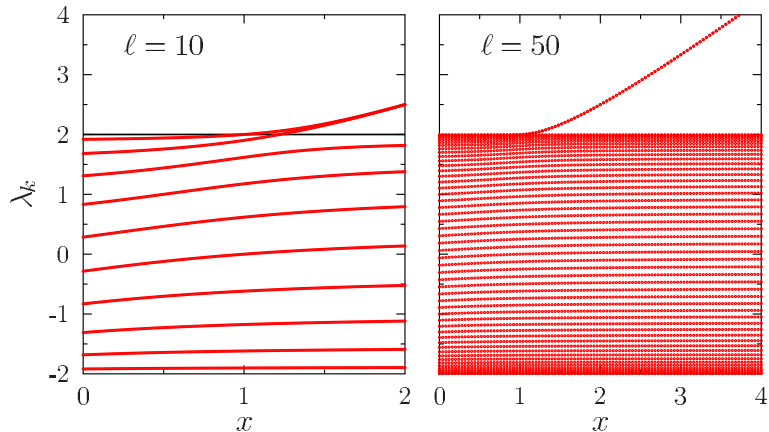}
\caption{Eigenvalues of the matrix~\eqref{e.trigmat1bound} for $y=1$
as a function of the corner element $x$, for matrix sizes $\ell=10$
and~$50$. When $x>1$ there can be two out-of-band eigenvalues.}
\label{f.fby1x}
\end{figure}

We remark that as long as there are no out-of-band eigenvalues
Eq.~\eqref{e.Parlett4} is exactly normalized, i.e.,
$\sum_k{O}_{k1}^2={1}$, a sum that in the large-$\ell$ limit turns
into the integral
\begin{equation}
 {\cal{I}}(x,y)= \int_0^\pi \frac{dk}\pi\,\frac{y^2\sin^2\!{k}}
 {[(2{-}y^2)\cos{k}-x]^2+y^4\sin^2\!{k}} = 1 ~.
\label{e.largellnorm}
\end{equation}
Eigenvalues $\lambda\notin[-2,2]$ can exist for large $x$ or $y$. Let
us consider the simpler case $y=1$, with $x>0$, for which numerical
results are shown in Fig.~\ref{f.fby1x}: it appears that when $x$ is
raised above 1 two eigenvalues can leave the band. Indeed, from
Eq.~\eqref{e.ximinori} one finds the secular equation
\[
 U_{\ell}-2xU_{\ell-1}+x^2U_{\ell-2}=0
\]
that by means of the representation~\eqref{e.Uch} is easily shown to
imply
\[
 x=\cosh{p}+\sinh{p}
 \Big[\tanh\frac{(\ell{-}1)p}2\Big]^{\pm1}
 ~~\ge~1 ~,
\]
and two out-of-band eigenvalues can indeed exist if $x>1$: in the
large-$\ell$ limit they converge to the same value
$\lambda=x+x^{-1}$. On the other hand, for $x<1$ all eigenvalues
belong to the band. This fact is reflected in the
integral~\eqref{e.largellnorm}, because ${\cal{I}}(x,1)=1$ for
$x\le{1}$, while ${\cal{I}}(x,1)=x^{-2}<1$ for $x>1$: indeed, the
full normalization requires the contribution from the out-of-band
components.

\begin{figure}
\centering
\includegraphics[width=100mm,angle=0]{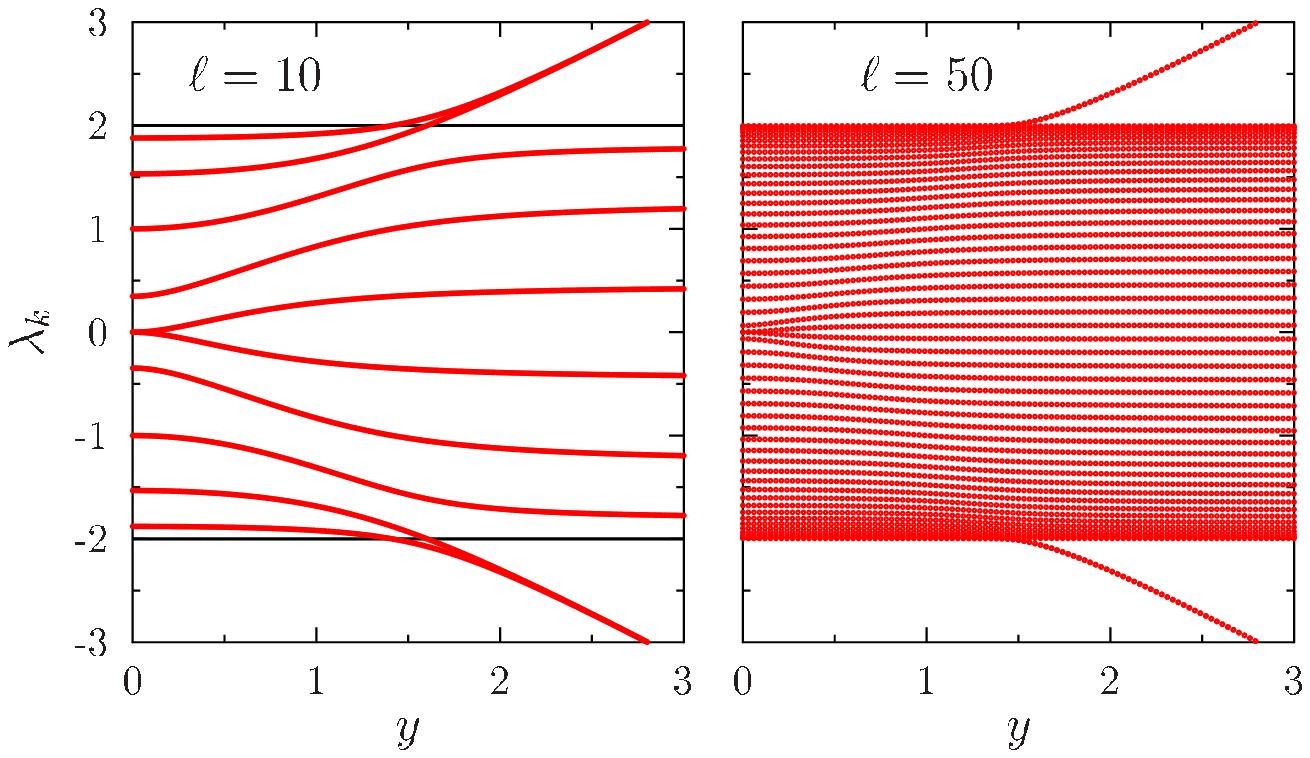}
\caption{Eigenvalues of the matrix~\eqref{e.trigmat1bound} for $x=0$
as a function of $y$, for matrix sizes $\ell=10$ and~$50$. When
$y>\sqrt{2}$ there can be two pairs of opposite out-of-band
eigenvalues.}
\label{f.fbx0y}
\end{figure}

A similar reasoning applies when $y$ is left to vary while $x=0$,
reported in Fig.\ref{f.fbx0y}, though in this case the out-of-band
eigenvalues occur as two pairs of opposite sign. We find indeed
${\cal{I}}(0,y)=1$ for $y\le\sqrt{2}$, while
${\cal{I}}(0,y)=(y^2{-}1)^{-1}<1$ for $y>\sqrt{2}$. It is to be noted
that calculating the integral~\eqref{e.largellnorm} in the general
case is not trivial; nevertheless, we can say that it evaluates to 1
as long as all eigenvalues belong to the band $[-2,2]$.

\subsection{More mirror symmetric elements}
As a second example we consider a mirror-symmetric matrix with more
nonuniform elements on the edges,
\begin{equation}\label{e.trigmatsenf}
\begin{pmatrix}
    0 & x \\
    x & 0 & y \\
    & y & 0 & 1 \\
    && 1 & 0 & 1 \\
    &&& \ddots &  \ddots &\ddots\\
    &&&& 1&  0 & 1 \\
    &&&&& 1&  0 & y \\
    &&&&&& y&  0 & x  \\
    &&&&&&& x &  0
\end{pmatrix}_{\ell\times\ell}~.
\end{equation}
Using straightforward algebra we find
\begin{align*}
  w_k & = \big[2-y^2-x^2 + (2-y^2) \cos{2k} + i y^2 \sin{2k}\big]^2~,\\
u(\xi) & = \big[2\xi^2(2-y^2)-x^2\big]^2-(1-\xi^2)\,
4 \xi^2y^4~, \\
t(\xi) & = 4\xi\,y^2\,\big[2\xi^2(2-y^2)-x^2\big]~, \\
\uu(\xi) & = \ul(\xi) =
\xi \big[-2 y^4 + x^2 (-2 + y^2) + 4 (2 - 2 y^2 + y^4) \xi^2\big]~, \\
\tu(\xi) & = \tl(\xi) =
y^2 \big[4 (2 - y^2) \xi^2 -x^2\big]~,
\end{align*}
and accordingly
\begin{align}
  \varphi_k &= 2k-
  \tan^{-1}\bigg[\frac{y^2 \sin{2k}}{z^2 + (2-y^2) \cos{2k}}\bigg]~,
\\
  {O}_{1 k}^2 = O_{k\ell}^2 &= \frac{2}{\ell{+}1{-}2\varphi_k'}
  ~\frac{x^2y^2\sin^2k}{\big[z^2 + (2-y^2) \cos{2k}\big]^2 + y^4 \sin^2{2k}}
\end{align}
where $z^2 \equiv 2 - x^2 - y^2$.

\subsection{Non-mirror-symmetric matrix}
\label{s.tridiagxy}

In order to connect our formalism with the results of
Ref.~\cite{Yueh2005} let us consider the following
non-mirror-symmetric matrix
\begin{equation}\label{e.trigmat2bscaled}
\begin{pmatrix}
    x & y &&&&&&  \\
    y & 0 & 1 &&&&&  \\
    & 1 & 0 & 1 &&&&  \\
    && 1 & 0 & 1 &&&  \\
    &&& 1 &  \ddots &\ddots&  \\
    &&&&\ddots&&1&\\
    &&&&& 1&  0 & 1  \\
    &&&&&& 1 &  z
\end{pmatrix}_{\ell\times\ell}~.
\end{equation}
where $\ell = n+2$. We find
\begin{align*}
\tu(\xi) & =  1~, &
\uu(\xi) & =  \xi-z~,
\\
 \tl(\xi) & =  y^2~,
 & \ul(\xi) & =  (2-y^2)\xi-x~,
\\
 t(\xi) & = 2\xi -x - y^2 z~,
 & u(\xi) & = (x - 2 \xi) (z - \xi) + y^2 (z \xi-1)~,
\end{align*}
and in particular
\begin{align}
  \tan 2\phi_k = \frac{(x + y^2 z - 2 \cos k) \sin k}{(x - 2 \cos k)
  (z - \cos k) + y^2 (z \cos k -1)}~,
  \label{e.yueh}
\end{align}
from which the spectral decomposition follows. In fact, it can be
shown that, once $O_{k1}$ is calculated with~Eq.\eqref{e.ok1cosk},
the remaining eigenvectors are given by~\eqref{e.wurstvec}, except
for the $\ell$th one that follows from Eq.~\eqref{e.evecrecur}.
Eq.~\eqref{e.yueh} extends the results of Ref.~\cite{Yueh2005}: for
example when $x=0$, $y=1$, and~$z=-1$ we find $2\phi_k=-\frac32k$ and
\[
 k_j = \frac{2\pi j}{2\ell+1}~,
\]
recovering Theorem 1 of Ref.~\cite{Yueh2005}. With the proper
parametrization it can be shown that the other theorems of
Ref.~\cite{Yueh2005} concerning symmetric tridiagonal matrices follow
as well.

\section{Conclusions}

We have introduced a technique for the analytical diagonalization of
large quasi-uniform tridiagonal (QUT) matrices. The quasi-uniformity
has been exploited to show that almost all eigenvalues belong to the
same band of those of the fully uniform matrix, $\lambda=2\cos{k}$,
with $k\in[0,\pi]$, and that their distribution is a deformation of
the equally spaced $k$'s of the uniform case, characterized by {\em
shifts} $\varphi_k$, as Eqs.~\eqref{e.phik} and~\eqref{e.kj} show.
The first components $O_{k1}$ of the normalized eigenvectors are
written in terms of ratios of low-degree
polynomials~\eqref{e.ok1cosk} that can be easily calculated from the
non-uniform part of the QUT matrix, while the other components are
constructed recursively from $O_{k1}$ using Eq.~\eqref{e.evecrecur};
exploiting the uniform-bulk property, i.e., using
Eq.~\eqref{e.solgen.Oknu}, all components can be expressed as
$O_{k1}$ times a combination of Chebyshev polynomials, as shown in a
particular example by Eq.~\eqref{e.Oknu}.

In the case of a large QUT matrix the eigenvalues can be described in
terms of a modified density of states within the band of the
corresponding uniform matrix. A limited number of out-of-band
eigenvalues can exist and have to be accounted for separately as
discussed in section~\ref{ss.oob} and exemplified in
section~\ref{ss.wurst}.

\section*{Acknowledgements}

The authors thank T.~J.~G.~Apollaro, A.~Cuccoli, and P.~Verrucchi for
fruitful discussions. L.~B. thanks M.~Allegra for helpfully reading
this manuscript.

\end{document}